\documentclass[aps, twocolumn, showpacs, superscriptaddress]{revtex4-1}
\usepackage{graphicx}
\usepackage{amsmath}
\usepackage{amsfonts}
\usepackage{amssymb}
\usepackage{hyperref}

\newcommand{\beq}{\begin{equation}}
\newcommand{\eeq}{\end{equation}}

\begin{document}

\title{Structural transitions of ion strings in quantum potentials}

\author{Cecilia Cormick} 
\affiliation{Theoretische Physik, Universit\"at des Saarlandes, D66123 Saarbr\"ucken, Germany} 

\author{Giovanna Morigi} 
\affiliation{Theoretische Physik, Universit\"at des Saarlandes, D66123 Saarbr\"ucken, Germany} 

\date{\today}

\begin{abstract}
We analyse the stability and dynamics of an ion chain confined inside a high-finesse optical resonator. When the dipolar transition of the ions strongly couples to one cavity mode, the mechanical effects of light modify the chain properties close to a structural transition. We focus on the linear chain close to the zigzag instability and show that linear and zigzag arrays are bistable for certain strengths of the laser pumping the cavity. For these regimes the chain is cooled into one of the configurations by cavity-enhanced photon scattering. The excitations of these structures mix photonic and vibrational fluctuations, which can be entangled at steady state. These features are signalled by Fano-like resonances in the spectrum of light at the cavity output. 
\end{abstract}


\maketitle

Crystals of singly-charged ions in traps are remarkable realisations of the phenomenon first predicted by Wigner \cite{Dubin_ONeil_RMP_1999}. The level of control experimentally achieved on these systems is impressive even at the quantum level and makes them promising candidates for several applications ranging from metrology to quantum information processing \cite{Haeffner_Roos_Blatt_PRep_2008,Wineland_RSL_2003,Schneider_Porras_Schaetz_2011}. Their versatility also offers the possibility to study paradigmatic models of strongly-correlated many-body systems \cite{Schneider_Porras_Schaetz_2011,Friedenauer_NatPhys_2008,Haeffner_Nat_2005,Timoney_Nat_2011,Kim_Nat_2010,Bollinger_JPB_2003}. 

Structural transitions in ion crystals have recently attracted renewed interest \cite{Fishman_PRB_2008,Retzker_PRL_2008,Gong_Lin_Duan_PRL_2010,DallaTorre_NatPhys_2010,Shimshoni_Morigi_Fishman_PRL_2011}. They are due to the interplay between the repulsive Coulomb interaction and the confining potential of Paul or Penning traps and can be controlled by varying, for instance, the aspect ratio of the trap potential. A prominent example is the linear-zigzag transition, that is classically described by Landau model \cite{Fishman_PRB_2008}, while its quantum analogue belongs to the universality class of a  one-dimensional ferromagnet \cite{Shimshoni_Morigi_Fishman_PRL_2011}. Studies of quenches across the instability in the classical regime showed that formation of defects follows the predictions of the Kibble-Zurek mechanism \cite{delCampo_PRL_2010,Zurek_PRep_1996}.

The combination of ion and dipolar traps \cite{Katori_Schlipf_Walther_PRL_1997, Schneider_NatPhot_2010} opens further perspectives, such as the possibility of realising the Frenkel-Kontorova model \cite{Garcia-Mata_Zhirov_Shepelyansky_2007,Pruttivarasin_NJP_2011}  and of coupling ultracold atomic systems with ions \cite{Idziaszek_Calarco_Zoller_PRA_2007, Zipkes_Nat_2010}. In Refs. \cite{Katori_Schlipf_Walther_PRL_1997, Schneider_NatPhot_2010}  the dipolar potential is classical, the quantum fluctuations of the electromagnetic field being very small. A very different scenario can be reached in presence of a cavity. Experiments with trapped ions in front of a mirror showed a mirror-mediated dipole-dipole interaction \cite{Eschner_Nat_2001} and demonstrated the mechanical effect of the vacuum state of the electromagnetic field on a single ion \cite{Bushev_PRL_2004}. The recent achievement of strong coupling between the optical transitions of ions forming a Wigner crystal and one mode of a high-finesse resonator \cite{Herskind_NatPhys_2009, Albert_PRA_2012} sets the stage for the observation of novel self-organised structures. In this regime, mechanical forces due to multiple scattering of a cavity photon can be infinitely long-ranged and may modify the structural stability even at the single photon level. The understanding of such dynamics can allow one to identify new control tools as well as to access new strongly-correlated states. The competition of long-range potentials of different nature, however, gives rise to a theoretical problem of considerable complexity. 

\begin{figure}[hbt]
\begin{center}
\includegraphics[width=0.36\textwidth]{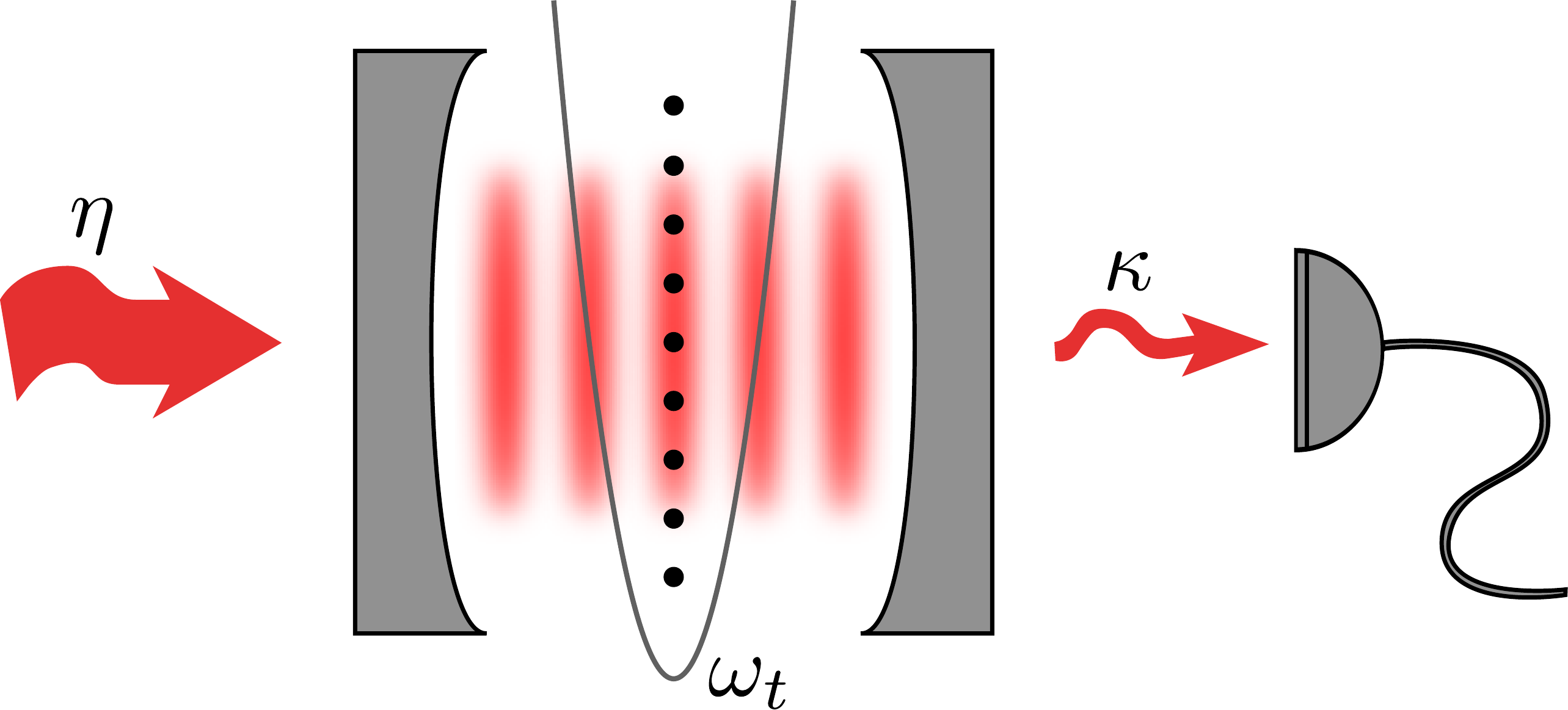}
\caption{\label{fig:system}The dipolar transitions of ions forming a chain couple with one mode of a high-finesse optical cavity. The depth of the optical potential inside the resonator depends on the ions' transverse positions and on the strength of the laser pumping the cavity. This property gives rise to hysteresis in the structural configuration and to quantum correlations between photonic and mechanical fluctuations. The scheme can be implemented in setups like the ones realised in \cite{Keller_Nat_2004,Stute_Nat_2012}.}
\end{center}
\end{figure}

In this Letter we theoretically characterise structural properties of crystalline structures inside a standing-wave resonator, analysing in particular how the crystal structure close to the linear-zigzag instability is modified in this environment.  Figure \ref{fig:system} displays the main features of the system. A string of $N$ ions of mass $m$ and charge $q$ is confined within an optical resonator by a radiofrequency trap, here described by a harmonic potential with axial and transverse frequencies $\omega_a$ and $\omega_t$, respectively. The dipolar transition of the ions interacts with a mode of the cavity field which is pumped by a laser with strength $\eta$. When cavity and pump are sufficiently out of resonance from the atomic dipole transition, the dynamical variables of ions and cavity field are described by the operators $\vec r_j$ and $\vec p_j$, denoting the position and momentum of the center of mass of the $j$-th ion in the array, and by the annihilation and creation operators $a$ and $a^{\dagger}$ of a cavity photon at frequency $\omega_c$. Their coherent dynamics are governed by Hamiltonian $H=H_{\rm cav}+H_{\rm ions}+H_{\rm int}$. Here, $H_{\rm cav} = -\hbar \Delta_c a^\dagger a  -{\rm i}\hbar \eta(a - a^\dagger)$ is the Hamiltonian for the cavity mode in absence of atoms and in the reference frame rotating at the pump frequency $\omega_p$ with $\Delta_c=\omega_p-\omega_c$, the Hamiltonian for the ions is given by
\begin{eqnarray}
&&\label{H:ions}
H_{\rm ions} = \!\sum_{j=1}^N \Bigg[\frac{{\vec p_j\,}^2}{2m} + V_{\rm trap} (\vec r_j) + \!\!\! \sum_{k=j+1}^N \!\!\! V_{\rm Coul} (|\vec r_j-\vec r_k|)\Bigg] ,\quad
\end{eqnarray}
and includes the kinetic energy, the trap potential $V_{\rm trap}$, and the Coloumb repulsion $V_{\rm Coul}$. Finally,  $H_{\rm int} = \hbar a^\dagger a U_0 (\vec r_1,\ldots ,\vec r_N)$ describes the interaction between ions and cavity field, with
\beq
\label{eq:U:0}
U_0(\vec r_1,\ldots ,\vec r_N)= \frac{g_0^2}{\Delta_0}\sum_j \cos^2(kx_j) e^{-y_j^2/\sigma^2}\,.
\eeq
Here, $g_0$ is the strength of the coupling between the cavity and the ions' transition at frequency $\omega_0$, $k$ the cavity wave vector, $\Delta_0=\omega_p-\omega_0$ the detuning of the pump from the dipolar transition, and $\sigma$ denotes the width of the cavity mode, that is generally smaller than the chain length: The number of ions coupling to the cavity mode, and hence contributing to $U_0$, is $N_{\rm eff}$, with $N_{\rm eff}<N$. Frequency $U_0$ weights the nonlinear coupling between motion and cavity mode: It is the shift of the cavity frequency due to the ions inside the resonator, and conversely it is the mechanical potential exerted on these ions by a single cavity photon \cite{Domokos_Ritsch_PRL_2002}. This term gives rise to mechanical effects that, for strong coupling, can be significant at the single-photon level. Incoherent effects arise from spontaneous decay of the dipolar transition at rate $\gamma$, cavity decay at rate $2\kappa$, and thermalization of the ions' motion with an external reservoir which may be due to patch potentials at the trap electrodes \cite{Haeffner_Roos_Blatt_PRep_2008, Wineland_NIST_1998}. We choose the detuning $|\Delta_0|$ to be the largest parameter, corresponding to the inequality $|\Delta_0|\gg\gamma,\kappa,|\Delta_c|,g_0\sqrt{\bar n}$, with $\bar n$ the mean intracavity photon number. In this regime the cavity-ion interaction is mainly dispersive and spontaneous emission can be neglected \cite{footnote:decay}.

The Hamiltonian $H$ formally agrees with the one derived for neutral atoms \cite{Maschler_Ritsch_PRL_2005, Domokos_Ritsch_PRL_2002}. However, while in \cite{Maschler_Ritsch_PRL_2005, Domokos_Ritsch_PRL_2002} the atomic interaction is a contact potential, here the ions repel via the long-range Coulomb repulsion. Therefore, in the first case the strength of the pump determines the quantum phase of the atoms in a non-trivial way \cite{Larson_PRL_2008}. For ions, on the other hand, quantum degeneracy is irrelevant but the strength of the cavity potential can substantially modify the crystalline structure. An effective dispersive potential for the particles can be derived when retardation effects can be discarded. In this limit the cavity field is determined by the instantaneous set of positions of {\it all} ions coupled with the cavity mode and reads  $\bar a=\eta/(\kappa-{\rm i}\Delta_{\rm eff})$, with $\Delta_{\rm eff}=\Delta_c-U_0$, while the corresponding effective potential takes the form 
\beq \label{eq:eff opt potential}
V_{\rm eff} = \left(\hbar \eta^2/\kappa \right)\arctan \left(-\Delta_{\rm eff}/\kappa \right)\,.
\eeq
This potential gives rise to an effective long-range force between these ions whose strength scales with the cooperativity $\mathcal C= g_0^2N_{\rm eff}/(\kappa |\Delta_0|)$. The ions' structure is then determined by the positions at which the total potential  $V_{\rm tot}=V_{\rm trap} + V_{\rm Coul} + V_{\rm eff}$ exhibits minima. 

\begin{figure}[hbt]
\begin{center}
\includegraphics[width=0.48\textwidth]{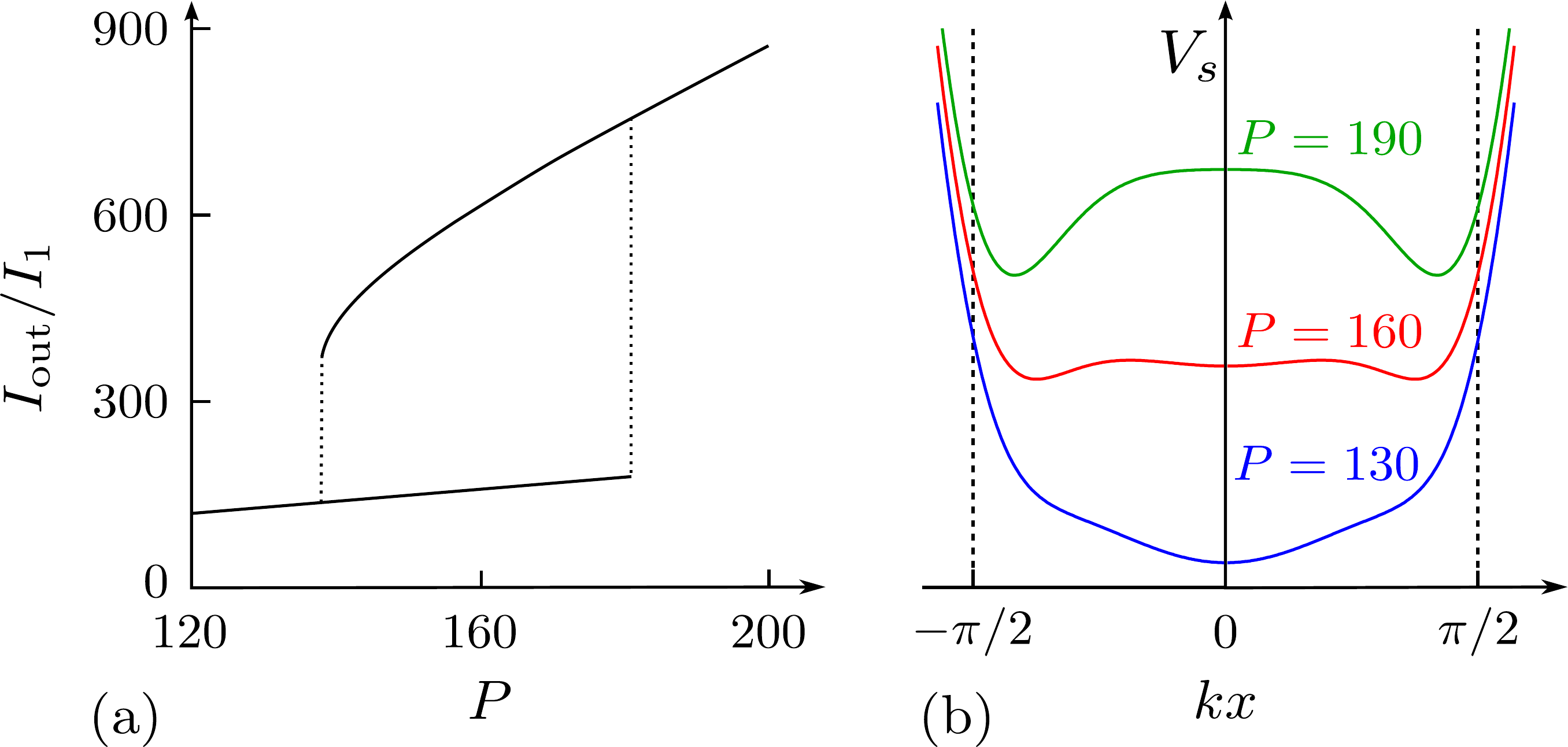}
\end{center}
\caption{\label{fig:hysteresis} (a) Intensity $I_{\rm out}$ at the cavity output for varying pump strength $P=( \eta/\eta_0)^2$, with $\eta_0^2=\kappa\omega_a^2/(2\omega_R)$. $I_{\rm out}$  is in units of $I_1 = I_{\rm out}(P=1)$ and is obtained for a configuration minimizing $V_{\rm tot}$, numerically found using linear and zigzag chains as initial guesses. (b) Potential $V_s$ as a function of the displacement $x$ of the central ion (in units of $1/k$) for $P=130,160,190$. The dashed lines indicate nodes at $kx=\pm\pi/2$. The plots correspond to a chain of 60 $^{40}$Ca$^+$ ions with interparticle distance 4.3 $\mu$m in the central region, coupled to a cavity mode with wavelength 866 nm and transverse width $\sigma=14\,\mu$m ($N_{\rm eff}\simeq 5.7$). The other parameters are $\omega_a/2\pi =$ 0.1 MHz, $\omega_t /2\pi= $ 2.26 MHz (the critical value is $\omega_{tc}/2\pi\simeq$ 2.216 MHz), $\Delta_c=0$, $\Delta_0/2\pi=$ 500 MHz,  $\kappa/2\pi=$ 0.5 MHz, $g_0/2\pi=$ 9.4 MHz, $\gamma/2\pi=$ 10 MHz, resulting in $\mathcal C = 2$.}
\end{figure}

Two situations can be identified depending on the value of the cooperativity $\mathcal C$. For $\mathcal C\ll 1$, the potential in Eq. (\ref{eq:eff opt potential}) approaches a classical potential whose depth is independent of the ions' positions. In this limit, when the ion string is parallel to the cavity axis (which corresponds to setting all values $y_j=0$), the system provides a realisation of the Frenkel-Kontorova model with trapped ions \cite{Garcia-Mata_Zhirov_Shepelyansky_2007}. When the string is instead orthogonal to the cavity-mode wave vector, as  in Fig. \ref{fig:system}, the optical potential generates a transverse force. This force is symmetric about the chain axis when the chain is at a node or antinode of the cavity standing wave. Then, close to the linear-zigzag mechanical instability the optical potential shifts the critical value of the transverse trap frequency with respect to the free-space value $\omega_{tc}$ \cite{footnote1}. In the following we shall assume that the equilibrium positions of the ions in the linear array are located at an antinode of the cavity standing wave. This can be realised, for instance, with the setups of Refs. \cite{Keller_Nat_2004,Stute_Nat_2012}. For blue-detuned pumps, with $\Delta_0>0$, the antinode is a maximum of the cavity potential and a mechanical instability thus appears at frequencies larger than $\omega_{tc}$, while a red-detuned pump field will have the opposite effect \cite{footnote2}. This behaviour is significantly modified at large cooperativities, $\mathcal C\gtrsim1$, where the cavity-mediated interaction between the ions becomes relevant. We consider $\Delta_c=0$, $\Delta_0>0$: In this case the intracavity field is minimum when the ions form a linear array, while it increases when their equilibrium positions arrange in a zigzag. This property can give rise to bistability of the linear and zigzag structures which can be observed in the mean value of the intensity $I_{\rm out}$ of the field at the cavity output. An example of this behaviour is shown in Fig. \ref{fig:hysteresis}(a) where $I_{\rm out}$ is plotted  as a function of the pump intensity $\eta^2$ for a chain of 60 ions in a harmonic trap, assuming that the central region of the chain couples to the cavity mode and $N_{\rm eff}\sim 5.7$. 

Further insight is gained by analysing the effective potential $V_s$ of the zigzag mode which is the soft mode of the linear-zigzag transition in free space \cite{Fishman_PRB_2008}. We first consider the simplest limit when the ions can be assumed to be equidistant (which describes the chain central region or a chain in an anharmonic axial trap \cite{Lin_EPL_2009}) and the mode amplitude reads $x_s = \sum_j (-1)^j x_j/\sqrt{N}$. Denoting by $\omega_s=\sqrt{\omega_t^2-\omega_{tc}^2}$ the frequency of the soft mode above the critical point in free space, the potential $V_s$ when the ions are uniformly illuminated by the cavity field and for $\omega_t>\omega_{tc}$ reads
\beq 
\label{V:s}
V_s = \frac{m}{2} \omega_s^2 x_s^2 + \frac{\hbar \eta^2}{\kappa} \arctan \left[\mathcal C \cos^2\left(\frac{kx_s}{\sqrt{N}}\right)\right]\,.
\eeq 
The second term on the right-hand side of Eq. \eqref{V:s} describes the effect of the optical field. For $\eta=0$ the cavity mode is in the vacuum state and the linear array is stable. The soft mode becomes unstable when the optical power is increased above the threshold value $\eta^2_{\rm th}=N(1+\mathcal C^2)/(4\mathcal C)\omega_s^2\kappa/\omega_R$, with $\omega_R=\hbar k^2/(2m)$ the recoil frequency. For $\mathcal C>1$ parameters can be found in which both linear and zigzag configurations are stable. In a finite chain the amplitude of the soft mode and the coupling of the ions to the cavity are not uniform along the chain. However, the potential energy as a function of the soft-mode amplitude gives similar qualitative results, as shown in Fig. \ref{fig:hysteresis}(b). Here, for certain values of $\eta$ the potential can exhibit three minima, corresponding to stable linear and zigzag arrays. We remark that the observed bistability is a consequence of the nonlinear dependence of the optical forces on the positions of the atoms within the standing-wave field. In the thermodynamic limit, if the region of the chain interacting with the cavity mode is finite, the effect of this coupling is a localized defect in the chain. For a finite system, nevertheless, forces acting on few ions can generate arrays close to zigzag configurations due to the long-range Coulomb repulsion \cite{Baltrusch_PRA_2011, Li_Lesanovsky_2012}.

The three metastable configurations can be observed when $\Delta_c=0$, $\Delta_0\gg\gamma$ as the result of a cooling process due the strong coupling with the cavity \cite{Vuletic_Chu_2000}. In this regime the excitations of the emergent crystalline structure reach a stationary state mixing photonic and vibrational modes. We analyse their behaviour by considering the coupled dynamics of the quantum fluctuations of field and motion. Be $\delta a=a-\bar a$ the quantum fluctuations of the field about the mean value $\bar a$, and $\{\delta x_j,\delta y_j\}$ the displacement of the ion localized at the equilibrium position $\vec r_j^{(0)}$ determined by the balance of harmonic, Coulomb and mean optical forces. For convenience we introduce the normal modes of the crystal, that characterise the dynamics of the ions when the coupling with the quantum fluctuations of the cavity field can be neglected. Let $\delta \zeta_j=\sum_n M_{jn}^{(\zeta)}B_{n0}(b_n+b_n^{\dagger})$ with $\zeta=x,y$ and $b_n$ ($b_n^\dagger$) the bosonic operator annihilating (creating) a phonon of the normal mode at frequency $\omega_n$, $B_{n0}=\sqrt{\hbar/(2m\omega_n)}$, and $M_{jn}^{(\zeta)}$ the element of the orthogonal matrix relating the local coordinates with the normal modes. The dynamics of normal modes and field fluctuations are governed by the Heisenberg-Langevin equations \cite{Gardiner_Collett_1985, Szirmai_PRA_2010}:
\begin{align}
\label{eq:a}
& \delta \dot a = (i \Delta_{\rm eff} - \kappa) \delta a - i \sum_n c_n \bar a (b_n + b_n^\dagger) + \sqrt{2\kappa}\, a_{\rm in}\,, \\
\label{eq:b}
& \dot b_n = -(i\omega_n + \Gamma_n) b_n -i c_n (\bar a^* \delta a + \bar a \delta a^\dagger) + \sqrt{2\Gamma_n} \, b_{{\rm in}, n}\,,
\end{align}
that include quantum noise on the cavity at rate $\kappa$ with corresponding input noise $a_{\rm in}$, and on the motion at rate $\Gamma_n$ with input noise $b_{{\rm in}, n}$, simulating the presence of a reservoir with which the ions' vibrations couple, such that $\langle b_{{\rm in}, n}^{\dagger}(t)b_{{\rm in}, m}(t')\rangle=N_n\delta_{n,m}\delta(t-t')$ \cite{footnote3}, with $N_n$ the mean phonon number at the temperature of the reservoir. Vibrations and field fluctuations couple with strength $c_n\bar a $, where
\beq
c_n = \frac{B_{n0}}{\Delta_0} \sum_j \left[M_{jn}^{(x)} \partial_{x_j} g_j^2 + M_{jn}^{(y)} \partial_{y_j} g_j^2\right] \label{eq:cavity-motion-coupling}
\eeq
and $g_j=g_0\cos (kx_j) {\rm e}^{-y_j^2/(2\sigma^2)}$. The coefficients $c_n$ vanish when all equilibrium positions are at field nodes, where $g_j=0$. If the particles are located at antinodes, the coupling is determined by the derivatives in $y$ direction which are assumed to be much smaller than those along $x$ ($k\sigma\gg1$). Thus, for the chosen setup the coupling between vibrations and field fluctuations is stronger in the zigzag configuration, while it is a very small perturbation when the ions form a linear chain. We remind that for the parameters considered the cavity field cools the normal modes coupled to it, so that cavity and crystal reach a stationary state \cite{Vuletic_Chu_2000}.

\begin{figure*}[hbt]
 \includegraphics[width=\textwidth]{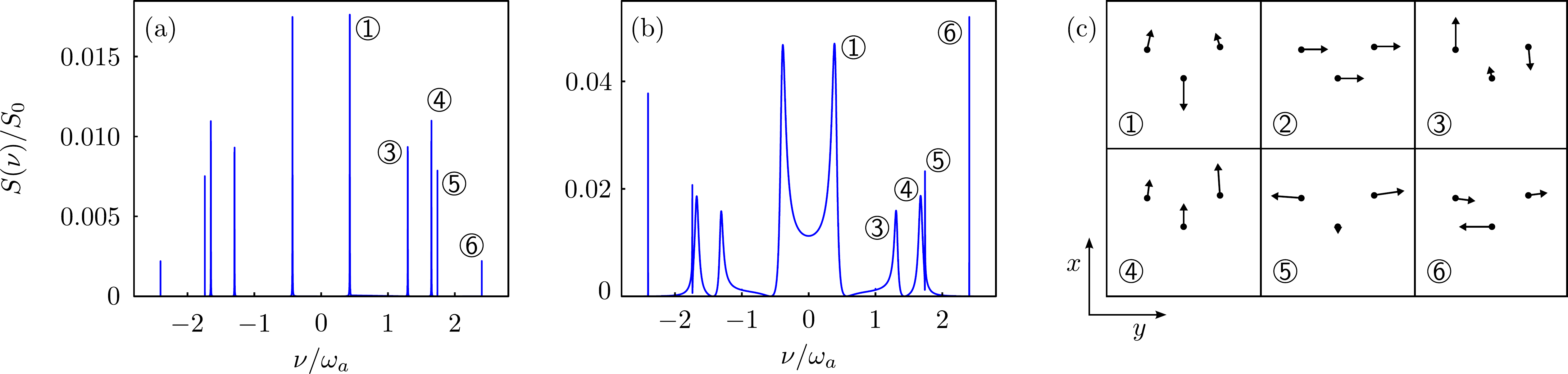}
 \caption{\label{fig:spectrum}Spectrum $S(\nu)$ of the field at the cavity output (in units of $S_0=1/\omega_a$) for a zigzag chain of 3 ions when only one ion at the chain edge, the right one in (c), couples significantly to the cavity mode (with Lamb-Dicke parameter $\sim$ 0.1) and the ions' motion is cooled by the cavity field. The parameters are the same as in Fig. \ref{fig:hysteresis}, expect for $\omega_a = \kappa = 2\pi~\times$ 1 MHz, $\omega_t = 2\pi~\times$ 1.57 MHz, and (a) $\mathcal C=0.5$, $P = 1$; (b) $\mathcal C=3$, $P = 0.22$; the mode width $\sigma$ is 0.65 times the interparticle distance in the linear array, so $N_{\rm eff}\simeq1.1$. The equilibrium positions are the same in (a) and (b). The motional modes contributing to the spectral peaks are sketched in panel (c) (not to scale; the resonance corresponding to mode 2 is not visible, because this mode is too weakly coupled to the cavity). The Rayleigh peak is not shown.}
\end{figure*}

We study the effect of this coupling in the spectrum at the cavity output, $S(\nu) = \langle a_{\rm out}(\nu)^\dagger ~ a_{\rm out} (\nu) \rangle/I_0$, with $I_0= 2 \kappa |\bar a|^2$ the zero-order intensity of the output field and $a_{\rm out} = a_{\rm in} + \sqrt{2\kappa} \, a$ \cite{Gardiner_Collett_1985}. The steady-state spectrum exhibits negligible fluctuations in the linear phase, while in the zigzag configuration it reads 
\begin{multline}
S(\nu)= S_0(\nu)
\Bigg\{ \frac{4\kappa |\theta(\nu)|^2 |\bar a|^2}{\kappa^2+(\nu-\Delta_{\rm eff})^2} \\
+ \sum_n c_n^2 \Gamma_n \left[ \frac{N_n}{\Gamma_n^2 + (\omega_n-\nu)^2} + \frac{N_n + 1}{\Gamma_n^2 + (\omega_n+\nu)^2} \right] \Bigg\}
\end{multline} 
where the first term is the contribution due to the coupling of the quantum vacuum with the crystal vibrations, with  $\theta(\nu) = \sum_n c_n^2 \omega_n/[\omega_n^2+(\gamma_n-i\nu)^2]$, and the second is due to thermal noise coupling to the modes. The common prefactor takes the form
\beq
S_0(\nu)=\frac{2}{\kappa^2 + (\nu+\Delta_{\rm eff})^2} \left|1+\frac{4\Delta_{\rm eff} \theta(\nu)|\bar a|^2}{(\kappa-i\nu)^2+\Delta_{\rm eff}^2}\right|^{-2} 
\eeq
and gives a Lorentz curve when ${\mathcal C}\ll 1$. Its functional behaviour is strongly modified when the cooperativity is increased: Then, motional and quantum noise do not simply add up, but nonlinearly mix to determine the spectral properties of the output field. 

Figure \ref{fig:spectrum} displays the spectra for a chain of three ions for different parameter choices: as $\mathcal C$ is increased the spectral lines change the relative heights, width, and shape. We first note the asymmetry in the spectra with respect to $\nu=0$: This is due to the (weak) coupling of the ions' motion to the thermal bath \cite{Wilson-Rae_NJP_2008}. The broadening at large cooperativity is a consequence of the vacuum input noise on the cavity field and indicates the rate at which the cavity cools the corresponding vibrational mode \cite{Bienert_2004}. It is accompanied by the appearance of Fano-like resonances which result from the dispersive effect of the cavity back-action and are a signature of quantum interference in the fluctuations of motion and field \cite{Fano_1961}. This interference is due to quantum correlations established by the dynamics described in Eqs. (\ref{eq:a})-(\ref{eq:b}), which can generate entanglement between vibrational and photonic modes \cite{Pirandola_PRA_2003}. In fact, for the parameters of Fig. \ref{fig:spectrum} (b) we find in the steady-state a logarithmic negativity  of 0.15 \cite{Vidal_Werner_PRA_2002} between cavity and phononic excitations. We remark that the field at the cavity output allows one to monitor the stationary state in a non-destructive way, it can be measured in existing experimental setups \cite{Dantan_PRL_2010} and could be used to realise feedback on the ion crystal, for instance by means of an appropriate generalization of the procedure in Ref. \cite{Bushev_PRL_2006}. 

In summary, the structural properties and quantum fluctuations of an ion Coulomb crystal inside a resonator are strongly affected by the nonlinearity of the cavity coupling. This effect is particularly visible close to structural instabilities. We have focused on the linear chain close to the zigzag instability and shown that for large cooperativity also the zigzag array can be made stable by the photon-mediated interaction between the ions (when a small region of the chain is coupled to the cavity, this coupling induces a localized defect in the chain with a zigzag form). This behaviour can be detected by hysteretical behaviour of the intensity of the field at the cavity output as a function of the pump strength. The excitations of these structures reach a stationary state where phonons and photons are strongly correlated and can exhibit entanglement. At even larger cooperativity the dynamics studied in this work could be induced by one photon inside the resonator, thereby providing an unprecedented control of many-body systems at the single-photon level. 

The authors thank I. Leroux, M. Drewsen, H. Ritsch, and S. Fishman for fruitful discussions, and F. Cartarius for help with numerical calculations. This work was partially supported by by the European Commission (STREP PICC, COST action IOTA, integrating project AQUTE), the Alexander von Humboldt and the German Research Foundations.


\begin{thebibliography}{30}
\bibitem{footnote:decay}
When the spontaneous decay is small but not negligible, for a given set of ions' positions the cavity decay rate reads $\kappa_{\rm eff}=\kappa + U_0 \gamma /(2\Delta_0)$ and also accounts for photon losses due to spontaneous emission.
\bibitem{footnote2}
The plane containing the zigzag array can be chosen to be such that micromotion is in the direction perpendicular to the plane and therefore has no effect on the problem under study.
\end{thebibliography}

\end{document}